\documentclass[prl,twocolumn, superscriptaddress, longbibliography]{revtex4}
\usepackage{lipsum}% http://ctan.org/pkg/lipsum
\usepackage{graphicx}% http://ctan.org/pkg/graphicx
\usepackage{float}
\usepackage{amsmath,amssymb,amstext,dsfont,tikz,graphicx,physics,mathtools, simpler-wick}
\usepackage[normalem]{ulem}
\usepackage[colorlinks,bookmarks=false,citecolor=blue,linkcolor=red,urlcolor=blue]{hyperref}

\begin{document}

\title{Entanglement transitions induced by quantum-data collection}
\author{Shane P. Kelly}
\email{skelly@physics.ucla.edu}
\affiliation{Mani L. Bhaumik Institute for Theoretical Physics, Department of Physics and Astronomy, University of California at Los Angeles, Los Angeles, CA 90095}
\affiliation{Institute for Physics, Johannes Gutenberg University of Mainz, D-55099 Mainz, Germany}
\author{Jamir Marino}
\affiliation{Institute for Physics, Johannes Gutenberg University of Mainz, D-55099 Mainz, Germany}

\date{\today}

\begin{abstract}
    We present an entanglement transition in an array of qubits, induced by the transfer of quantum information from a system to a quantum computer.
    This quantum-data collection is an essential protocol in quantum machine learning algorithms that promise exponential advantage over their classical counterparts. 
    In this and an accompanying work, we identify sufficient conditions for an entanglement transition to occur in the late time state of the system and quantum computer.
    In this letter, we present an example entanglement transition occurring in a system comprised of a 1D chain of qubits evolving under a random brickwork circuit.
    After each layer, a fraction $p$ of sites undergo noisy quantum transduction in which quantum information is transferred to a quantum computer but at the cost of introducing noise from an environment.
    For an entanglement transition to occur, we argue that the environment must obtain the same amount of information as gained by the computer.
    Under this condition, the circuit shows a transition from volume law to area law entanglement as the rate $p$ is increased above a critical threshold.
    Our work delineates the prerequisites for quantum-data collection to induce entanglement transitions, thereby establishing a foundational framework for emergent entanglement phenomena in protocols relevant to quantum machine learning.
\end{abstract}

\maketitle

The transfer of information is a ubiquitous process in any experiment, technological development or daily interaction.
Most of the time, these events involve simple conversion processes, and a direct analysis of channel capacities, or sampling of entropy, is no longer necessary in their design.
Yet, as experiments become more elaborate, a direct analysis of channel capacity and information flow can become advantageous.
This occurs in the classical realm, when machine learning is required to analyze large data sets~\cite{mackaylearning}, but it is also important in quantum experiments where sensing beyond the standard quantum limits~\cite{RevModPhys.89.035002,Toth_2014, RevModPhys.90.035005} requires the observation of several quantum devices prepared in entangled states~\cite{RevModPhys.90.025004,franke2023quantum,doi:10.1126/science.abi5226,PhysRevLett.113.103004}.

A key difference in the quantum domain is that quantum information cannot be copied~\cite{wootters1982single,dieks1982communication}.
Thus, the interrogation of a quantum system requires the transfer of information between the system and measurement apparatus, and necessarily results in disrupting the natural information dynamics of the system.
This is sharply demonstrated by the measurement induced phase transition~(MIPT)~\cite{fisherreview,PhysRevB.99.224307,PhysRevX.9.031009,Li_2018} that can occur in a unitary many body quantum system interrogated by a projective measurement process~\cite{PRXQuantum.2.010352,PhysRevB.102.014315,Lavasani_2021,PhysRevResearch.3.023200,PhysRevLett.127.235701,10.21468/SciPostPhysCore.5.2.023,PhysRevB.104.155111,Bao_2021,PhysRevLett.129.080501,PhysRevLett.126.170602,PhysRevX.11.041004,PhysRevLett.128.010605,theoryOftransitionsBao,PhysRevB.101.104302,PhysRevB.105.064305,Sierant2022dissipativefloquet,PhysRevD.106.046015,PhysRevLett.127.140601,MIPTdp1,PhysRevLett.130.220404,PhysRevB.106.214303,PhysRevLett.130.120402,PhysRevB.107.L220201,PhysRevB.108.075151,PhysRevA.108.022426,akhtar2023measurementinduced,PhysRevB.108.L020306,PhysRevB.107.214203,Viotti2023geometricphases,tikhanovskaya2023universality,PhysRevLett.130.120402,murciano2023measurementaltered,wang2023entanglement,PhysRevLett.129.080501,poboiko2023measurementinduced,poboiko2023theory,10.21468/SciPostPhys.7.2.024,PhysRevLett.126.123604,Biella2021manybodyquantumzeno,PhysRevB.103.224210,PhysRevB.105.L241114,liu2024entanglement,liu2024noiseinduced}.
At a large measurement rate, the projective measurements disrupt information spreading, while below a critical threshold, quantum information spreads and generates strongly entangled states.

In this setting, classical information, in the form of projective measurement outcomes, is transferred to the measurement apparatus.
More recently, the transfer of quantum information has been shown to achieve exponential advantage in machine learning tasks compared to the transfer of classical information~\cite{quantumadvantage,boundsonlearning,aharonov2022quantum}.  
The advantage is gained in a quantum-enhanced experiment~\cite{quantumadvantage} where the measurement apparatus doesn't perform a projective measurement, but instead transfers quantum states to a collection of qubits for post-processing by a quantum computer.
This quantum-data collection~\cite{Cerezo_2022} may seriously disrupt the natural dynamics of information in the system. 
This motivates us to investigate the possibility that the disruption results in a phase transition, similar to the MIPT.

Previous work~\cite{weinstein2022scrambling} observed that the perfect transfer of quantum information does not result in an entanglement transition similar to the MIPT.  
In the MIPT, the transition occurs at a critical measurement rate, below which volume law entangled states are produced and above which area law entangled states are produced.  
In the previous work~\cite{weinstein2022scrambling}, projective measurements were replaced by a swap operation that transferred the complete qubit state to a measurement apparatus.  
While a different transition was observed, the entanglement transition was lost: The late-time system states were short-range entangled mixed states independent of the swap rate. 
Thus, it appears that the transfer of classical information is somehow distinct from the transfer of quantum information.

In the work accompanying this Letter~\cite{long}, we argue the difference is not fundamental, and systematically investigate how to resolve it. 
The difference is that projective measurements naturally possess a symmetry in the information dynamics that the swaps do not.
The symmetry, which we call the information exchange (IE) symmetry, is presented below and results in an entanglement transition when spontaneously broken~\cite{long}.
Since swaps of a single qubit break this symmetry explicitly, the entanglement transition does not occur in protocols akin to the one discussed in Ref.~\cite{weinstein2022scrambling}.

\begin{figure}[t]
	\includegraphics[width=\columnwidth]{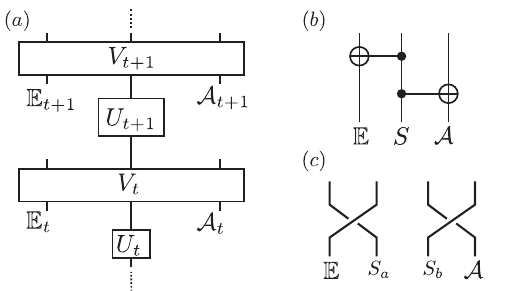}
        \caption{(a) Dynamics of an experiment in which information is transferred from a system $S$ to an apparatus $\mathcal{A}$.
When the apparatus $\mathcal{A}$ couples to the system, noise is generally introduced and is modelled by a Markovian environment $\mathbb{E}$ with $T$ uncorrelated components $\mathbb{E}_t$ for $t=1\dots T$. The unitary $V_t$ models the coupling operation which produces the signal in the $t^{th}$ component of the apparatus $\mathcal{A}_t$ and injects noise due to the environment $\mathbb{E}_t$. (b) Unitary implementation of a projective measurement on a single qubit. (c) Noisy-transduction operation acting on the system of two qubits $S_a$ and $S_b$. This operation stores one qubit $S_b$ in the quantum apparatus, while the other $S_a$ is erased and discarded into the environment.}
	\label{fig1}
\end{figure}

Nonetheless, the collection of quantum information does not explicitly forbid the IE symmetry.
Thus, there may exist operations that posses this symmetry, collect quantum-information and induce an entanglement transition.
In this Letter, we demonstrate that this is indeed the case.
Below, we present the simplest operation, shown in Fig.~\ref{fig1}(c), that possesses this symmetry and transfers quantum information to a quantum apparatus.

In contrast to projective measurements, an environment must be explicitly introduced, and it is no longer clear if the quantum-conditional entropies considered in the MIPT remain entanglement quantifiers.  
Surprisingly, we find that the IE symmetry also guarantees the quantum-conditional entropies are quantifiers of entanglement.
As a result, we show that the transition, tuned by the rate of quantum information transfer, is again between area law and volume law entangled states.

\textbf{A framework for experiments:} 
We consider experiments of the general form shown in Fig.~\ref{fig1}(a) in which a system of interest undergoes unitary dynamics and is periodically coupled to a measurement apparatus.
We allow the coupling to the measurement apparatus to introduce noise and therefore introduce explicitly an environment Hilbert space, $\mathbb{E}$.
We also allow quantum information to be transferred to the measurement apparatus, and therefore model the apparatus with an additional Hilbert space $\mathcal{A}$.

The experiments are comprised of $T$  two-part steps. 
In the first part, the system evolves under its natural unitary dynamics $U(t)$.
In the second part, a measurement process occurs in which the system is coupled to the $t^{th}$ component of the measurement apparatus, $\mathcal{A}_t$.
The measurement process introduces Markovian noise, and this is modeled by an environment $\mathbb{E}$, also with $T$ independent components which we label $\mathbb{E}_t$ for $t=1\dots T$.
The measurement process is described by a unitary $V(t)$ acting between the system and the $t^{th}$ components of the apparatus and environment.
The full dynamics of the system, environment and apparatus is shown in Fig~\ref{fig1} and is described by a unitary $\mathcal{U}_T=\prod_{t=1}^T V(t)U(t)$.

In this Letter we assume the environment, system, and measurement apparatus are initialized in a pure state $\ket{\psi_\mathbb{E}}\otimes\ket{\psi_0}\otimes\ket{\psi_{\mathcal{A}}}$ with no correlations between the $T$ different components of the apparatus and environment.
This ensures the $T$ measurement processes are independent.
We treat more general initial conditions in the Supplemental Material~(SM).

\begin{figure*}
  \centering
  \includegraphics[width=1.\textwidth]{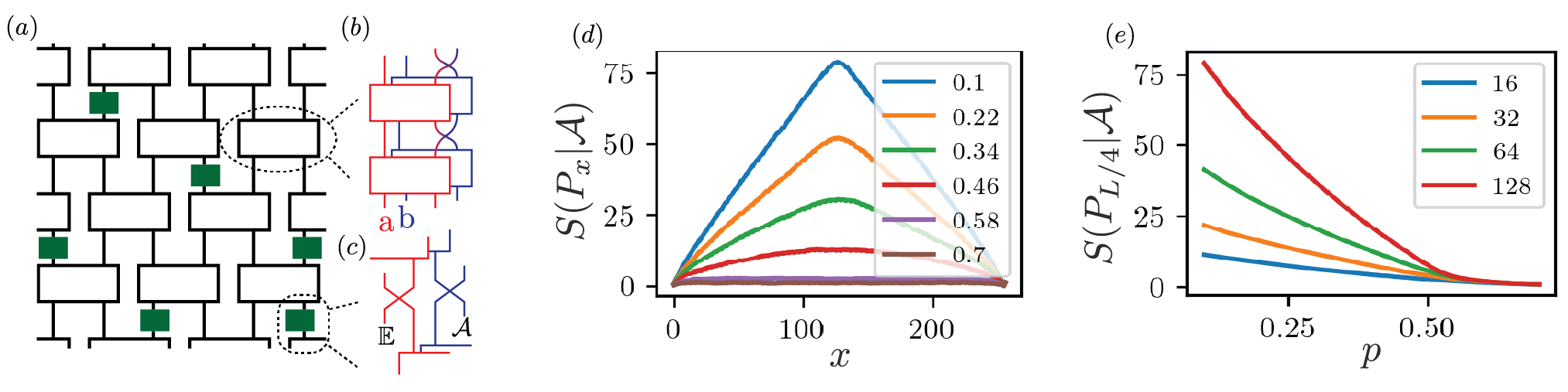}
  \caption{(a) Brickwork circuit showing an entanglement transition tuned by the rate of noisy-transduction operations~(shown in green). Qubits a and b are grouped into one line. (b) Four two-qubit unitaries composing a brick. The unitaries acting on the a qubits are shown in red, while those acting on the b qubits are shown in blue. (c) The noisy-transduction operations swaps a qubits into the environment, and b qubits into the apparatus. Both apparatus and environment qubits are initialized in pure states.  (d) Volume law and area law scaling of quantum-conditional entropy $S(P_x|\mathcal{A})$ for a continuous subsystem composed of $x$ sites, and conditioned on the state in the apparatus. Here we consider a system of $L=128$ sites; the legend gives the corresponding noisy-transduction rate $p$. (e) Quantum conditional entropy $S(P_x|\mathcal{A})$ for $x=L/4$ as a function of $p$; the legend shows the number of sites $L$. A transition from volume law to area law scaling appears at $p=0.52$. A careful scaling analysis is performed in the SM.}
  \label{fig3}
\end{figure*}

While not obvious, it is known that unitary dynamics of this form can model projective measurements~\cite{theoryOftransitionsBao,nielsen2010quantum,RevModPhys.75.715}.
The unitary coupling implementing a projective measurement on a single qubit~\cite{theoryOftransitionsBao,nielsen2010quantum} is shown in Fig.~\ref{fig1} (b).
There, a controlled-not (CNOT) gate is first applied between the measured system qubit and a qubit in the apparatus.
Then, a second CNOT gate is applied to the system qubit and a qubit in the environment.
If the initial state of the system qubit is $\ket{\psi}=\psi_0\ket{0}+\psi_1\ket{1}$  and the initial state of the apparatus is $\ket{0}$, the mixed state on the system and apparatus, after the two CNOT gates, will be $\rho = \left|\psi_0\right|^2 \ket{00}\bra{00}+\left|\psi_1\right|^2 \ket{11}\bra{11}$.
This statistical distribution of quantum states captures the uncertainty of the measurement outcome.
By conditioning on one of the outcomes in the apparatus, for example $0$, the state in the system will be the pure state $\ket{0}$ obtained by applying the corresponding projector.

Here we emphasize that projective measurements are symmetric under the exchange of the apparatus and environment, see Fig.~\ref{fig1} (b).
This guarantees that the apparatus gains the same amount of information as the environment.
This equivalence is the essence of the IE symmetry that is spontaneously broken in the MIPT~\cite{long} and which we present below.

To ensure the IE symmetry in a quantum-data-collection protocol, we introduce an operation that transfers an equal amount of quantum information to the apparatus and to the environment.
The operation, which we refer to as noisy transduction, is shown in Fig.~\ref{fig1}(c).
This operation acts on two qubits of the system, $S_a$ and $S_b$.
The state of qubit $S_b$ is transduced~\cite{Lauk_2020} into the apparatus by a swap gate applied between it and a purified qubit in the apparatus.
While the quantum state of another qubit $S_a$ is destroyed by a swap gate applied between it and a purified qubit in the environment.
At the end of the process, the environment qubits are discarded, while the apparatus qubits are stored for quantum postprocessing.

The essential difference between projective measurements and noisy transduction is that quantum information can be acquired by the latter.
For the projective measurements, the final state in the apparatus is incoherent in the measurement basis. 
For the noisy-transduction operation, coherence is allowed in any arbitrary basis, allowing for arbitrary quantum states to be collected.

\textbf{The quantum-data collection experiment:} This noisy-transduction operation can induce an entanglement transition. 
Similar to the MIPT, the experiment showing the transition comprises a brickwork random circuit acting on $L$ sites, with $U(t)$ implementing either the even or odd layers of the brickwork. 
In the MIPT circuit, the sites contains a single qubit and $V(t)$ implements projective measurements.
In contrast, the quantum-data-collection experiment contains two qubits per site, labeled a and b, and $V(t)$ implements the noisy-transduction operation on each site with probability $p$.
If a site is transduced, then qubit a is transferred to the environment and qubit b is transferred to the apparatus, see Fig.~\ref{fig3}.

To guarantee the IE symmetry for the full circuit, we require the system unitary $U_t$ to be symmetric under exchange of the a and b qubits.
This is accomplished as follows. 
At each step, an identical unitary chosen from the Clifford group, or Haar measure, is applied to both the $a$ and $b$ qudits of two neighboring sites.  
Next the two sets of $a$ and $b$ qubits are swapped with each other.
Finally, another pair of identical two-qubit random unitaries are applied. 
The full sequence of unitary and swaps comprising a brick is shown in Fig.~\ref{fig3}(b).

Finally, we note that the choice of a brickwork structure and random unitaries is simply for analytic and numerical control of the problem. 
Similar to other random unitary approaches~\cite{fisherreview}, the results are expected to hold for any chaotic unitary dynamics, so long as they are symmetric under the exchange of the $a$ and $b$ qubits. This symmetry ensures that an equal amount of quantum information is gained by the apparatus as is lost to the environment. 
As argued in the accompanying work~\cite{long}, this symmetry in information is required for replica symmetry breaking to occur as it does in the MIPT~\cite{theoryOftransitionsBao}.

\textbf{The entanglement transition:} 
The information and entanglement dynamics of this experiment are similar to the MIPT.
The phenomenology for both experiments is captured by the quantum-conditional entropy~\cite{PhysRevLett.79.5194,PhysRevA.60.893} $S(A|B)=S(\rho_{AB})-S(\rho_{B})$, defined using the von Neumann entropy $S(\rho)=-\tr[\rho\ln\rho]$ on two subsystems $A$ and $B$, where $\rho_A=\tr_B[\rho_{AB}]$.
For convenience, we refer simply write $S(A)=S(\rho_A)$ such that the conditional information is expressed as $S(A|B)=S(A\cup B)-S(B)$.

The entanglement transition is observed in $S(P_S|A)$, where $P_S$ is an arbitrary subset of the system qubits.
Below the critical rate $p<p_c$, the quantum-conditional entropy~$S(P_S|A)$ scales with the volume of the region $P_S$, while above the critical rate $S(P_S|A)$ obeys an area law and is independent of the size of $P_S$.
This is shown in Fig.~\ref{fig3}, where we present results from numerical simulation~\cite{gottesman_stabilizer_1997,gottesman_heisenberg_1998,aaronson_improved_2004}.
There the unitaries are chosen randomly from the two-qubit Clifford group for system sizes up to $L=256$.
In Fig.~\ref{fig3}(d) we show the quantum-conditional entropy $S(P_x|\mathcal{A})$ for a contiguous set of $\left|P_x\right|=x$ qubits.
The dependence on the size of the region $P_x$ shows both area law scaling for $p<0.514$ and volume law scaling for $p>0.514$.
To demonstrate the transition between these two behaviors, we show the quarter-cut conditional entropy in Fig.~\ref{fig3}(e), as a function of $p$.

In the SM, we also consider analogues of purification dynamics~\cite{Gullans_2020,PhysRevLett.125.030505}, and use the tripartite mutual information~\cite{PhysRevB.101.060301} to identify the critical point at $p=0.514$ and critical exponent of $\nu=1.16$ distinct from the analogous exponent in MIPT~($\nu\approx 1.3$)~\cite{Gullans_2020,PhysRevB.101.060301, PhysRevB.100.134306}. 
A distinct universality class from the MIPT is also observed for unitaries chosen from the Haar measure~\cite{long}. 

\textbf{Entanglement structure:} 
Since the state on the system and apparatus is mixed, the von Neumman entropy is not an entanglement quantifier~\cite{RevModPhys.81.865}.
Nonetheless, the above transition is an entanglement transition.
To demonstrate this, we introduce the information exchange~(IE) symmetry.
In Ref.~\cite{long} we show that this symmetry is spontaneously broken both in the MIPT and in the above circuit.
Here we use the IE symmetry to determine the entanglement structure of the mixed state describing the system and apparatus qubits.

\begin{figure}[t]
  \centering
      \includegraphics[width=\columnwidth]{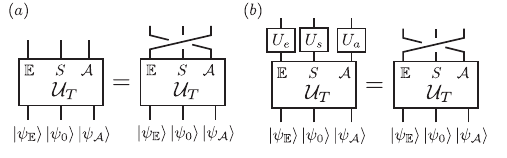}
      \caption{Two symmetries that an experiment collecting either quantum or classical data can possess: (a) apparatus-environment exchange symmetry (AEE), which requires the experiment to be symmetric under exchange of apparatus and environment, and (b) the Local Unitary Exchange symmetry~(LUE) is shown. It also requires that the experiment be symmetric under the exchange of the apparatus and environment, but only up to some unitary transform $U_e\otimes U_s\otimes U_a$ which is separable across environment, system, and apparatus.}
  \label{fig2}
\end{figure}

The IE symmetry is the weakest form of three similar symmetries.
For all three symmetries, the transformation is the exchange of the apparatus and environment. 
The three symmetries are distinguished by what is required to be symmetric.
The strongest requirement is for the Apparatus-Environment Exchange~(AEE) symmetry, which requires the full quantum state to be symmetric, see Fig.~\ref{fig2}(a).
Projective measurements satisfy this symmetry since the two CNOT gates in Fig.~\ref{fig1}(b) commute.

We found that the state equivalence required by the AEE symmetry is unnecessary to guarantee an entanglement transition.
Instead, it is sufficient to require equivalence up to a local unitary applied independently of the system, apparatus and environment, see Fig.~\ref{fig2}(b).
We call this weaker form of symmetry the Local Unitary Exchange (LUE) symmetry.
While the AEE symmetry doesn't hold for the noisy-transduction operation, the LUE symmetry does.
In this case, the local unitary on the system is the exchange of the two system qubits $a$ and $b$,  and the local unitaries on the apparatus and environment are the identity.

The LUE symmetry guarantees that the entropies satisfy 
\begin{eqnarray}
    S(P_S\cup\mathcal{A})=S(P_S\cup\mathbb{E}),
\end{eqnarray}
which hold for any $P_S$.
This implies that an observer of the apparatus and a fictitious observer of the environment have equivalent information about the system, $S(P_S|\mathcal{A})=S(P_S|\mathbb{E})$. 
We designate any experiment that satisfies this relation for all $P_S$ and times, $T$, as satisfying the IE symmetry.
The IE symmetry is the weakest of the three symmetries, and only requires that, after the exchange of the apparatus and environment, the entropies $S(P_S\cup\mathcal{A})$ are unchanged for all $P_S$.

The IE symmetry guarantees that the regions $P_S$ and $P_S^c\cup \mathcal{A}$ are entangled.
Here we have taken $P_S^c$ to be the complement of $P_S$ within the system such that $P_S\cup P_S^c = S$.
This is shown as follows.
First, the IE symmetry, along with the purity of the global state $\ket{\psi_T}$ implies $S(P_S|\mathcal{A})=-S(P_S|P_S^c\cup\mathcal{A})$.
Second, the results of Refs~\cite{horodeckiPartialQuantumInformation2005,horodeckiQuantumStateMerging2007}, show that a negative quantum-conditional entropy of magnitude $\left|S(A|B)\right|$ implies the ability to distill $\left|S(A|B)\right|$ Bell pairs between the two regions.\footnote{
While the negative quantum-conditional entropy gives the minimum number of distillable Bell pairs, it is not an entanglement monotone because it can be both positive and negative and is not strictly zero for unentangled states.}
Thus, the area law to volume law transition in $S(P_S|\mathcal{A})$ is equivalent to a transition in the number of distillable Bell pairs between $P_S$ and $P_S^c\cup A$

The same symmetry and arguments hold for the entanglement phenomenon in the MIPT except that the state on the apparatus is diagonal in the measurement basis.
The nature of classical measurements therefore forces all entanglement to be on the system (between $P_S$ and $P_S^c$).
In contrast, entanglement can also spread onto the apparatus when quantum information is collected.
Finally, subadditivity can be used to show~\cite{SM} that the IE symmetry implies $S(P_S|\mathcal{A})\geq0$ which guarantees finite entanglement in both the area law and the volume law phase.

\textbf{Discussion:} 
In this work we have demonstrated that quantum data collection can induce an entanglement transition.
The transition generalizes the MIPT and is also induced by the backaction of the data collection protocol on the unitary dynamics.
The transition was demonstrated numerically for Clifford gates, and in Ref.~\cite{long} the transition is established for gates chosen from the Haar measure. 

There we also establish the generality of entanglement transitions induced by data collection, quantum or classical.
A replica approach is used, and we show how the LUE and IE symmetries guarantee a minimal subgroup of the replica symmetry.
When this minimal subgroup is spontaneously broken, the generalized form of the MIPT transition results.
Crucially, we demonstrate that these transitions spontaneously break the IE symmetry.

This allowed us to easily identify the noise-transduction protocol that possess this symmetry and transfers quantum information to the apparatus, all without having to derive a replica theory.
Thus, while the specific implementation above is easily simulated numerically, there are many other experiments that possess the IE symmetry.
An example is a collection of interacting atoms with an inversion symmetry that emit photons in opposite directions:
Information contained in photons moving in one direction~(say, towards the apparatus) is equal to the information moving in the opposite direction~(the environment).

Future work might consider the effect of breaking the IE symmetry systematically.
Progress in this direction has been made by introducing an environment to the circuits with projective measurements~\cite{PhysRevLett.129.080501,liu2024entanglement,liu2024noiseinduced,lovas2023quantumcodingtransitionspresence}.
One might consider breaking the IE symmetry in the opposite way and introducing an apparatus to record an excess of quantum information.

Furthermore, many aspects of criticality should be considered, such as determining the conformal theory governing the critical point, and the effect of dimensionality.
It is likely that free fermion or bosonic models will aid in this discussion, as was the case for the MIPT~\cite{Buchhold_free_fermions_PRL,Buchhold_long_ranged,Buchold_free_fermions_PRX,Giulia_Russomanno_issing,Hafezi_continuous_monitoring,OG_monitored_fermions,Poboiko_Mirlin_replica_theory,poboiko2023measurementinduced,schiro_measured_subradiance,szyniszewski_lunt_arijeet_disordered_fermions,Turkeshi_2022_stocastic_reseting,Turkeshi_Fazio_Free_fermion_zero_clicks,Turkeshi_Piroli_neagtivity_fermions}.
Furthermore, cross-entropy benchmarks and shadow tomography techniques adapted for observing the MIPT~\cite{crossentropyOG,crossEntropyZ2sym,garratt_shadows} will need to be reconsidered now that the apparatus contains quantum states.

Our work also provides a framework to understand more elaborate measurement operations.
For example, instead of coupling the IE symmetry to the exchange of the a and b qubits, it might be coupled to a range of other symmetries.
Furthermore, it would be interesting to consider generalization of the IE symmetries in which more than two parties~(environment and apparatus here) compete over information on the system.
Finally, the IE symmetry and its symmetry breaking determine bounds on quantum communication tasks~\cite{SM}, and it may yield insight into gaining advantage in quantum machine learning~\cite{Cerezo_2022,boundsonlearning,quantumadvantage}.

\textbf{Acknowledgements:}
We greatly appreciate comments by the referees, Vince Hou, Zack Weinstein and Gerald E. Fux. We thank Martino Stefanini, Dominic Gribben, Oksana Chelpanova, and Riccardo Javier Valencia Tortora for valuable discussions on the information exchange symmetry. We are also grateful to Zack Weinstein and Ehud Altman for stimulating ideas.
We acknowledge support by the Deutsche Forschungsgemeinschaft (DFG, German Research Foundation) – Project-ID 429529648 – TRR 306 QuCoLiMa (“Quantum Cooperativity of Light and Matter”), by the Dynamics and Topology Centre funded by the State of Rhineland Palatinate, and by the QuantERA II Programme that has received funding from the European Union’s Horizon 2020 research and innovation programme under Grant Agreement No 101017733 (’QuSiED’) and by the DFG (project number 499037529).
The authors gratefully acknowledge the computing time granted on the supercomputer MOGON 2 at Johannes Gutenberg-University Mainz (hpc.uni-mainz.de).

\bibliographystyle{apsrev4-1}
\bibliography{refs}
\end{document}

\title{Supplemental material for: Entanglement transitions induced by quantum data collection}
\author{Shane P. Kelly}
\email{skelly@physics.ucla.edu}
\affiliation{Mani L. Bhaumik Institute for Theoretical Physics, Department of Physics and Astronomy, University of California at Los Angeles, Los Angeles, CA 90095}
\affiliation{Institute for Physics, Johannes Gutenberg University of Mainz, D-55099 Mainz, Germany}
\author{Jamir Marino}
\affiliation{Institute for Physics, Johannes Gutenberg University of Mainz, D-55099 Mainz, Germany}

\date{\today}

\maketitle

\tableofcontents

\section{Framework for considering temporal correlations}

In the main text we introduced a brickwork quantum circuit $\mathcal{U}_T$ that couples system, apparatus, and environment.
We considered the system, apparatus, and environment initialized in a pure state $\ket{\psi_{T=0}}=\ket{\psi_\mathbb{E}}\otimes\ket{\psi_0}\otimes\ket{\psi_{\mathcal{A}}}$ and discussed the spacial correlations of the late time state $\ket{\psi_{T}}=\mathcal{U}_T\ket{\psi_{T=0}}$, occurring after $T=4L$ layers of the brickwork circuit.
In particular, we quantified the spacial correlations via the von Neumann entropies
\begin{eqnarray}
    S(P_S; M)\equiv S(\tr_{P_S^c\cup M^{c}}[\ket{\psi_T}\bra{\psi_T}])
\end{eqnarray}
for the reduced density matrix of $\ket{\psi_T}$, on the subsystems $P_S\cup M$, where $P_S$ is a subset of the system, and $M\subset \mathcal{A}\cup \mathbb{E}$ is a subset of the apparatus $\mathcal{A}$ and environment $\mathbb{E}$.

As mentioned in the main text the entanglement transition also manifests in the temporal correlations between the initial state and final state.
To capture these correlations, we initialize the system $S$ in a maximally entangled state $\ket{\text{Bell}(S,S')}$ with an auxiliary copy of the system $S'$.
Both $S$ and $S'$ contain $2L$ qubits respectively.
As the system $S$, environment $\mathbb{E}$, and apparatus $\mathcal{A}$ evolve under the unitary $\mathcal{U}_T$, the auxiliary system $S'$ remains unchanged and acts as a reference for the initial state of $S'$.
The final evolved state is 
\begin{eqnarray}
    \ket{\tilde{\psi}_{T}}=\mathcal{U}_T\ket{\psi_\mathbb{E}}\otimes\ket{\text{Bell}(S,S')}\otimes\ket{\psi_{\mathcal{A}}},
\end{eqnarray}
and we will consider the von Neumann entropies 
\begin{eqnarray}
    S(P_S,P_S'; M)\equiv S(\tr_{P_S^c\cup P_S'^c\cup M^{c}}[\ket{\tilde{\psi}_T}\bra{\tilde{\psi}_T}])
\end{eqnarray}
to capture temporal correlations between the initial and final state.
Note that by discarding the auxiliary qubits $S'$, the initial state becomes a maximally mixed state.
In this scenario, $S(P_S,\varnothing;M)$ captures the late time correlations.
For clarity, we use the symbol $S_{\psi}(P_S;M)$ for the scenario in which the initial state of the system is in a pure state $\ket{\psi_0}$.

\section{Entanglement properties of IE symmetric experiments}
The IE symmetry guarantees a few important properties of the conditional entropies which relate $S(P_S|\mathcal{A})$ and $S(P_S,P_S'|\mathcal{A})$ to a quantification of entanglement. 
The first property is the complement symmetry
\begin{eqnarray}~\label{eq:c}
    S(P_S,P_{S'};\mathcal{A}) = S(P_S^c,P_{S'}^c;\mathcal{A}),
\end{eqnarray}
which follows from the IE symmetry and the purity of the global state $\ket{\mathcal{U}}$.

The second property is due to subadditivity applied to the environment, apparatus, and the system
\begin{eqnarray}
    S(\varnothing,\varnothing;\mathcal{A})+
    S(\varnothing,\varnothing;\mathbb{E}) \leq
    S(P_S,P_S';\mathcal{A})+
    S(P_S,P_S';\mathbb{E}),
\end{eqnarray}
which in combination with the IE symmetry yields the positivity of the quantum conditional entropy
\begin{eqnarray}
    S(P_S,P_S'|\mathcal{A})\geq 0.
\end{eqnarray}
The same arguments hold for $S(P_S|M)$ such that $S(P_S|\mathcal{A})\geq 0$

Finally, by using the IE symmetry and complement symmetry, we can obtain
\begin{eqnarray*}
    S(P_S,P_S'|\mathcal{A})=-S(P_S\cup P_S'|\mathcal{A}\cup P_S^c\cup P_S'^c),
\end{eqnarray*}
which implies that, in the volume law phase, the quantum conditional entropy $S(P_S\cup P_S'|\mathcal{A}\cup P_S^c\cup P_S'^c)$ is negative and scaling with volume.
As discussed in the main text, negativity of quantum conditional entropy implies entanglement between the regions $P_S\cup P_S'$ and $P_S^c\cup P_S'^c\cup \mathcal{A}$.
For unitary dynamics interspersed with projective measurements, the state in the apparatus is classical, and the entanglement is simply between the system regions as expected.

\section{Critical exponents of the brickwork quantum data collection circuit}

We now present numerical results for the coherent information and the entanglement dynamics for the brickwork circuit comprised of two-qubit unitaries chosen from the Clifford group.
Instead of preforming complex algorithms~\cite{Koenig_2014} to uniformly sample the Clifford group, we sample the two-qudit unitaries in the following way.
First, we uniformly apply one of the six single-qubit Clifford unitaries to each of the two qubits with equal probability.
Then, with a probability $0.9$ we apply a CNOT followed again by two random single qubit unitary applied to each qubit.
With probability $0.1$ we apply only the random single-qubit unitaries.
While this procedure doesn't sample all possible two-qubit Clifford gates, we believe it is sufficiently generic to capture the universal features of the entanglement transition.

\begin{figure}[t]
	\includegraphics[width=\columnwidth]{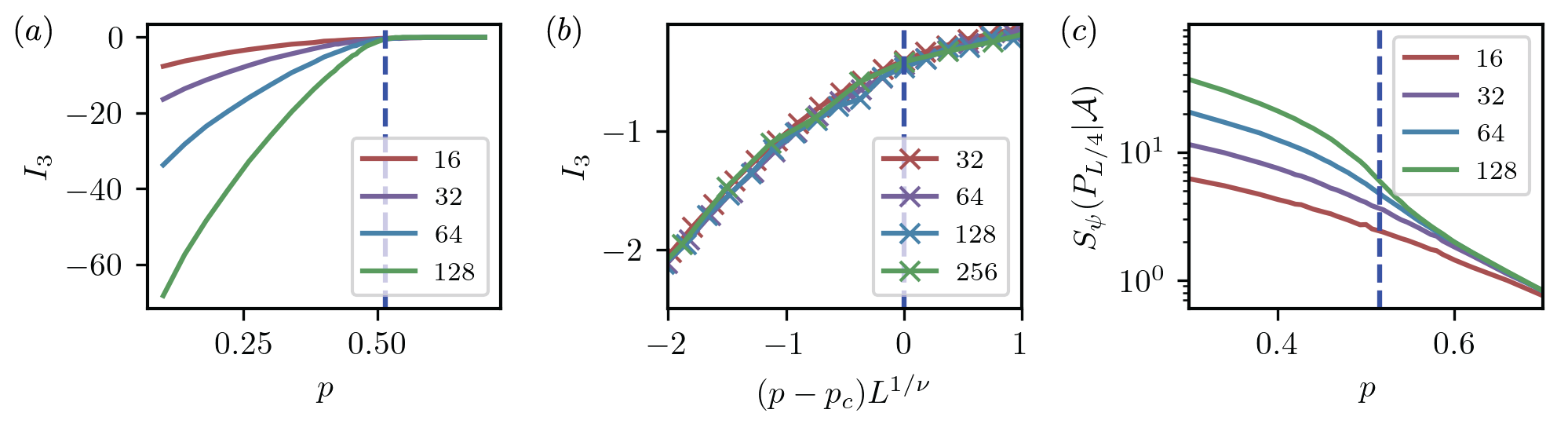}
\caption{Information-exchange symmetry breaking after $T=4L$ layers of the brickwork circuit for the system initialized in a pure product state.  The number of sites $L$ is shown in the legend of each figure, and the critical probability $p_c=0.514(2)$ is shown by a blue dashed vertical line. (a) Conditional tripartite mutual information defined in Eq.~\ref{eq:tp} v.s. $p$. (b) Scaling collapse of tripartite mutual information for $\nu=1.16(3)$; (c) Quarter-cut quantum-conditional entropy $S_{\psi}(P_{L/4}|\mathcal{A})$ showing logarithmic scaling close to $p=p_c$.}
	\label{fig:spacial}
\end{figure}

In the main text we showed evidence that this circuit exhibits a IE-symmetry-breaking phase transition by considering the final state quantum-conditional entropy $S_\psi(P_x|\mathcal{A})$, where $P_x$ included the $a$ and $b$ qubits from site $1$ to $x$.
To determine the critical point, we consider the tripartite mutual information, which was the most successful probe for identifying critical exponents of the measurement induced phase transition~(MIPT)~\cite{PhysRevB.101.060301}.
In a system with periodic boundary conditions, which is translationally invariant after disorder averaging, the tripartite mutual information, conditioned on the apparatus is constructed as
\begin{eqnarray}\label{eq:tp}
    I_3=4S_\psi(R_1|\mathcal{A})-2S_\psi(R_1\cup R_2|\mathcal{A})-S_\psi(R_1\cup R_3|\mathcal{A})
\end{eqnarray}
where $R_n$ contains the qubits at sites between $(n-1)L/4$ to $nL/4$.
$I_3$ is useful for identifying critical properties because of how it scales. 
Above the critical noisy transduction rate, $S_\psi(P_x|\mathcal{A})$ obeys an area law (independent of $x$) and the conditional tripartite information vanishes $I_3=0$.
While below the critical noisy transduction rate, $S_\psi(P_x|\mathcal{A})$ scales linearly with $x$ such that $I_3$ is negative and scales linearly with system size.
The conditional tripartite mutual information is particularly useful if at the critical point $S_\psi(P_x|\mathcal{A})$ scales logarithmically in $x$.
In this case, a direct analysis of $S_\psi(P_x|\mathcal{A})$ is difficult owing to a difficulty in distinguishing logarithmic from small power law scaling. 
Instead, the conditional tripartite mutual information is a universal constant when $S_\psi(P_x|\mathcal{A})$ scales logarithmic.

In Fig.~\ref{fig:spacial}, we see that the tripartite mutual information shows all three behaviors.
Below the critical noisy transduction probability $p<p_c$, we find $I_3$ is negative and scaling with system size.
Above criticality $p>p_c$ we find $I_3\rightarrow 0$, and at criticality we find it is equal to a constant $I_3=-0.5$ independent of system size.
This suggests logarithmic scaling of $S_\psi(P_x|\mathcal{A})$, which we confirm qualitatively in Fig.~\ref{fig:spacial}(c).
Since $I_3$ is constant at the critical point, identifying the critical point $p_c=0.514(2)$ is straightforward.
Furthermore, we find scaling collapse when rescaling probability around the critical point by $(p-p_c)L^{1/\nu}$, where $\nu=1.16(3)$.

\section{Dynamical purification and quantum communication transitions}

\begin{figure}[t]
        \includegraphics[width=0.7\columnwidth]{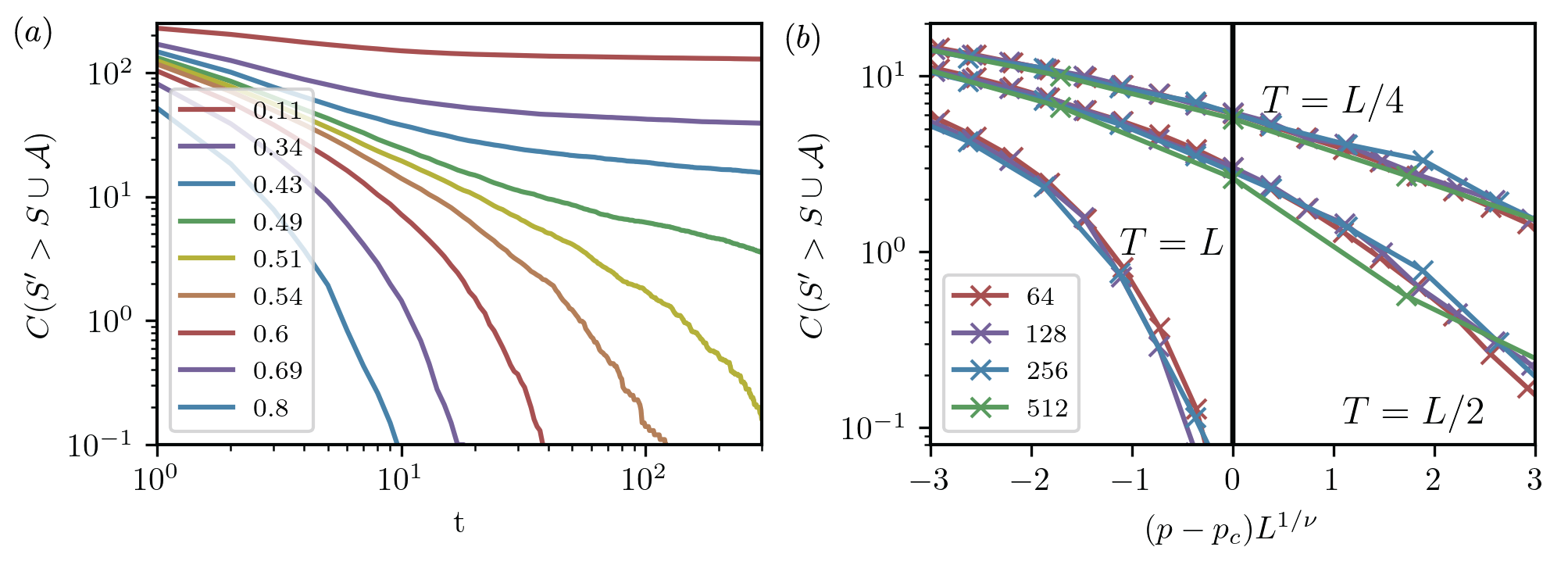}
\caption{Information exchange symmetry breaking shown by the quantum channel capacity between the initial state of the system, and the final state of both the system and apparatus. The channel capacity is quantified by the coherent information $C(S'>S\cup \mathcal{A})$. (a) Coherent information v.s. time for a set of noisy transduction rates shown in the legend. In this figure, the system size is $L=128$. (b) Scaling collapse for the coherent information close to the critical noisy transduction rate $p_c$. The legend gives the system size $L$. No fitting was preformed here. Instead, the critical parameters $p_c=0.514$ and $\nu=1.16$ used are the optimal values obtained for the conditional tripartite mutual information $I_3$. Finally, we note that the number of random samples considered ranged from $32$, for $L=512$, to $24000$ for $L=64$. The number of samples for $L=512$ is extremely limited owing to each site containing $2$ qubits such that the simulations are preformed on $1028$ qubits.}
	\label{fig:temporal}
\end{figure}

In the MIPT, it was recognized early on that the entanglement transition can be viewed as a dynamical purification transition~\cite{Gullans_2020}, or a transition in quantum coding~\cite{PhysRevB.103.104306,PhysRevB.103.174309,PhysRevLett.125.030505,PhysRevLett.125.070606}.
Similarly, we show in Ref.~\cite{long} that the information exchange symmetry guarantees a similar transition in the quantum-channel capacity of the channel between the initial system state and the finial state of the system and apparatus.
This channel capacity is given by the coherent information
\begin{eqnarray}
    C(S'>S\cup\mathcal{A})=S(S,\varnothing;\mathcal{A})-S(S,S';\mathcal{A})
\end{eqnarray}
which gives the maximum number of qubits of information encoded in the initial state $S'$ that can be recovered from the finial state of the system and apparatus $S\cup\mathcal{A}$.
Numerical Clifford simulations of this quantity are shown in Fig.~\ref{fig:temporal} and show a transition at the same critical point $p=0.514$ with the same exponent $\nu$ as for the quantum conditional entropy discussed above.

In the MIPT the channel capacity~(the coherent information) between $S'$ and $S\cup\mathcal{A}$ was related to the purification dynamics of an initially maximally mixed system.
A similar relation holds in general, and follows from the IE symmetry $S(S,S';\mathcal{A})= S(\varnothing,\varnothing;\mathcal{A})$ such that
\begin{eqnarray}
    C(S'>S\cup\mathcal{A})=S(S,\varnothing;\mathcal{A})-S(S,S';\mathcal{A})=S(S,\varnothing;\mathcal{A})-S(\varnothing,\varnothing;\mathcal{A})=S(S,\varnothing|\mathcal{A})
\end{eqnarray}
Thus, the phase transition can be observed by monitoring the dynamics of both the quantum-conditional entropy $S(S,\varnothing|\mathcal{A})$ and the coherent information $C(S'>S\cup\mathcal{A})$.
Subadditivity also guarantees that $C(S'>S\cup\mathcal{A}) =S(S,\varnothing|\mathcal{A})\geq0$.

Finally, the IE symmetry guarantees that the channel between initial system $S'$ to finial system $S$ and environment $\mathbb{E}$ shows the same transition, $C(S'>S\cup\mathcal{A})=C(S'>S\cup\mathbb{E})$.
Thus, the IE symmetry guarantees the existence of a symmetry between two tasks: when decoding using the apparatus is possible, it is also possible with the environment.
In Ref.~\cite{long} we argue that this transition can be viewed as a spontaneous symmetry breaking of this symmetry between two tasks.

\section{Efficient simulation}
To simulate the expectation value of any observable on both system and apparatus, the numerical simulation would require keeping track of $N=O(L^2)$ qubits because each time step introduces $O(L)$ apparatus qubits and the simulation evolves for $O(L)$ time steps.
For stabilizer states, the memory requirement to represent such a state scales as $O(N^2)$, such that simulating $O(L^2)$ qubits requires $O(L^4)$ memory and would not currently be feasible for the system sizes we considered.
Fortunately, the entropies observing the transition are efficiently simulated by the method discussed below.

\subsection{Review of stabilizer states}
To circumvent the $O(L^4)$ complexity, we make use of the structure of stabilizer states to compute $S(P_S,P_S';\mathcal{A})$ with only $O(L^2)$ memory complexity.
A stabilizer state on $N$ qubits has the form
\begin{eqnarray}
    \rho=\prod_{i=1}^{M}\frac{(1+g_i)}{2}
\end{eqnarray}
where $g_i$ is a set of $N$ commuting operators of the form
\begin{eqnarray}\label{eq:gi}
    g_i=\gamma_i\prod_{x=1}^{N}X_i^{\alpha_{i,x}}z_i^{\beta_{i,x}}
\end{eqnarray}
where $X_i$ and $Z_i$ are Pauli operators, and the vectors $\vec{\alpha}_i$ and $\vec{\beta}_i$ are 2 vectors with components $\alpha_{i,x}\in\{0,1\}$.
Up to the phase $\gamma_i\in\{1,-1,i,-i\}$, the operators $g_i$ are equivalent to a vector $\vec{v}_i=\vec{\alpha}_i \oplus \vec{\beta}_i$ living in the $2N$ dimensional vector space with discrete field $\mathcal{F}_2$~(additional and multiplication defined module $2$).
The operators $g_i$ generate the commuting stabilizer group under operator multiplication.
The representation of $g_i$ in the $2N$ dimension vector space defines an isomorphism between the stabilizer group and an $M$ dimensional subspace, where the stabilizer group multiplication maps to vector addition.
A stabilizer state is then equivalently represented by either a group presentation of the stabilizer group, or by $M$ linear independent vectors which span the $M$ dimensional subspace.
The second is used in numerical simulations, and in the worse case, a pure state requires $M=N$ vectors in a $2N$ dimensional vectors require memory scaling as $O(N^2)$.

The von Neumann entropy of a stabilizer state is $S(\rho)=N-M$, and to compute, $S(P_S,P_S';\mathcal{A})$, we simply need to identify the number of generators for the stabilizer group of the reduced state on the subsystem $P_S\cup P_S' \cup \mathcal{A}$.
The inefficient algorithm accomplishes this by tracking the evolution of the full stabilizer state on $S\cup S' \cup \mathcal{A}$, and computing the reduced state on $P_S\cup P_S' \cup \mathcal{A}$ when needed.
Before introducing the efficient algorithm, we review the necessary ingredients for the inefficient algorithm: preforming unitary evolution, and applying the depolarization map on a subset of qubits $B$~(computing the reduced state on the qubits $B^c$).
Unitary evolution follows from $U\rho U^{\dagger}=\prod_i (1+U g_i U^{\dagger})/2$, and for Clifford unitaries~(which preserve the structure Eq.~\ref{eq:gi}), the unitary corresponds to a linear map on the $M$ dimensional subspace.
The depolarization of $K$ qubits can also be viewed as a projective linear map which maps the $M$ dimensional subspace of a $2N$ dimensional vector space to an $M'\leq M$ dimensional subspace of a $2(N-K)$ dimensional vector space.

These linear maps can be implemented efficiently by careful consideration of the $2K$ dimensional subspace describing the support of the generators $g_i$ on the $K$ qubits.
For example, a unitary acting on only $K$ qubits can be implemented by a linear transformation on the $2K$ dimensional subspace.
Depolarization of $K<N$ qubits is less efficient and is preformed in the following steps:
\begin{enumerate}
    \item Gaussian elimination.
    \item Discarding generators with support on the depolarized qubits.
    \item Discarding the trivial support of the remaining generators on the depolarized qubits.
\end{enumerate}
The second and third steps are trivial, while the first step corresponds to a change of basis of the $M$ dimensional subspace, $\vec{v}_i\rightarrow \vec{\tilde{v}}_i$, such that 
\begin{enumerate}
    \item only $N_p\leq 2K$ transformed vectors, $\vec{\tilde{v}}_i$, have nonzero projection, $P_K\vec{\tilde{v}}_i\neq 0$ to the $2K$ dimension subspace describing support on the $K$ qubits.
    \item The $N_p$ vectors have linearly independent projections on to the $2K$ dimensional subspace.
\end{enumerate}
The second step discards the $N_p$ vectors, and the third step applies the projection $1-P_K$ to the remaining vectors.
Afterwards the reduced state is represented as an $M'=M-N_p$ dimensional subspace of a $2(N-K)$ dimensional vector space.
With these two linear maps, the dynamics of the system coupled to the apparatus and environment can be simulated with $O(L^4)$ memory complexity.

\subsection{The efficient algorithm}

An efficient algorithm is constructed by realizing the unitary dynamics on the apparatus qubits are trivial after they have coupled to the system.
This means the state of an apparatus qubit does not affect the system after it has coupled, and the reaming unitary dynamics can be computed without knowing the support of $g_i$ on the apparatus.
Similarly, the depolarization process only requires knowing the projections of $v_i$ on the qubits being depolarized.
The main difference with a unitary operation, is the Gaussian elimination $\vec{v}_i\rightarrow \vec{\tilde{v}}_i$ effects the projection of the vectors onto the apparatus subspace.
Still, one does not need to track this effect because it has no later effect on projection of the vectors within the system subspace.
Finally, computing the entropy of a reduced state only requires knowing the dimensionality of the stabilizer subspace and not the specific form of it.
Thus, we can compute the dynamics of the entropies $S(P_S,P_S';\mathcal{A})$ without tracking the support of the generators on the apparatus qubits that have already coupled to the system.
The main limitation of doing so is that we can not compute observables with support on the apparatus or entropies $S(P_S,P_S';M)$ where $M\neq \mathcal{A}$.

Furthermore, since the apparatus qubits that have not yet coupled to the system are in a trivial product state, we need not keep track of their state explicitly.
We therefore only introduce them at the time they interact with the system.
Explicitly, at the beginning of the dynamics, a stabilizer pure state is initialized on $S\cup S'$ described by a $4L$ dimension subspace of a $8L$ dimension vector space on $\mathcal{F}_2$.
As a reminder, each system site contains $2$ qubits, such that $S$ and $S'$ contain $4L$ qubits.
Numerically, we initialize the state with two completely mixed auxiliary qubits which we use to compute the effect of noisy transduction.
As the dynamics evolve, unitary gates transform the support of $\vec{v}_i$ on the system, which we track explicitly.

We preform noisy transduction as follows.
First, we initialize the two auxiliary qubits in the initial state of the environment and apparatus.
In Hilbert space this looks like
\begin{eqnarray}
    \rho\rightarrow \rho\otimes \rho_a\otimes \rho_e=\prod_{i=1}^{M}\frac{(1+g_i)}{2}\frac{(1+g_a)(1+g_e)}{4}
\end{eqnarray}
if states $\rho_a$ and $\rho_e$ are initially pure, and 
\begin{eqnarray}
    \rho\rightarrow \rho\otimes \rho_a\otimes \rho_e=\prod_{i=1}^{M}\frac{(1+g_i)}{2}
\end{eqnarray}
if they are maximally mixed.
Numerically, additional vector $\vec{v}_a$ and $\vec{v}_e$ representing $g_a$ and $g_e$ are added to the stabilizer subspace for the pure state, while nothing is done for a maximally mixed qubit.
The swap gates are then applied, and the auxiliary qubit representing the environment qubit is depolarized, but by skipping step 3 to keep the auxiliary qubit in a maximally mixed state for future use.
Then, since we are neglecting the support on the apparatus, the vectors $\vec{v}_i$ are projected to the system subspace, trivializing their support on the auxiliary qubits.

Without any further steps, the number of generators still grows in time as the apparatus qubits are added to the state.
This means, the memory complexity would grow as $O(L^2T)$ as we track the support of $O(LT)$ vectors on an $O(L)$ dimensional subspace.
This again is inefficient as we do not gain anything by tracking the structure of the linearly dependent sets of vectors.
Therefore, after each step in the brickwork, we apply Gaussian elimination on the $S\cup S'$ subspaces.
This will result in a set of vectors having trivial support on the system. 
We do not need to keep track of these vectors explicitly as they transform trivially under unitary and depolarization channels on the system qubits.
Still, we cannot ignore them completely since they must have nontrivial support on the apparatus and contribute to the dimensionality of the stabilizer subspace.
Thus, in addition to the nontrivial vectors, we keep track of the number of trivial vectors we ignore after this Gaussian elimination.
Due to this procedure, we only need to keep track the number of ignored vectors, and at most $8L$ vectors describing the nontrivial support of the stabilizer subspace on the system.
Thus, our memory requirements scale only as $O(L^2)$ allowing for the simulations presented above.

\bibliography{suppref}